\documentclass[sigconf,nonacm]{acmart}
\usepackage{popets}

\setcopyright{popets}
\copyrightyear{YYYY}

\acmYear{YYYY}
\acmVolume{YYYY}
\acmNumber{X}
\acmDOI{XXXXXXX.XXXXXXX}
\acmISBN{}
\acmConference{Proceedings on Privacy Enhancing Technologies}
\usepackage{algorithmic}
\usepackage{graphicx}
\usepackage{stfloats}
\usepackage{textcomp}
\usepackage{xcolor}
\usepackage{graphicx}
\usepackage{caption}
\usepackage{subcaption}  
\def\BibTeX{{\rm B\kern-.05em{\sc i\kern-.025em b}\kern-.08em
    T\kern-.1667em\lower.7ex\hbox{E}\kern-.125emX}}

\usepackage[normalem]{ulem}
\usepackage{rotating}
\usepackage{tablefootnote}
\usepackage{tabularray}
\usepackage{hyperref}
\usepackage{microtype}
\usepackage{graphicx}
\usepackage{subcaption} 
\usepackage{booktabs} 

\usepackage{hyperref}

\usepackage{mathtools}
\usepackage{amsthm}

\usepackage[capitalize,noabbrev]{cleveref}

\theoremstyle{plain}

\theoremstyle{definition}

\theoremstyle{remark}

\usepackage[textsize=tiny]{todonotes}

\usepackage{colortbl} 

\usepackage{multirow}

\usepackage{algorithm}      
\usepackage{caption}
\usepackage{subcaption}
\usepackage{graphicx}
\usepackage{caption}
\usepackage{subcaption}
\usepackage{caption}
\usepackage{subcaption}

\begin{document}

\title[LibFHE: A Numba-Based CUDA-Python Library for Non-RNS CKKS-BGV Fully Homomorphic Encryption on GPUs]{LibFHE: A Numba-Based CUDA-Python Library for Non-RNS CKKS-BGV Fully Homomorphic Encryption on GPUs}

\author{John Chiang}
\orcid{0000-0003-0378-0607}
\email{john.chiang.smith@gmail.com}

\renewcommand{\shortauthors}{John Chiang}

\begin{abstract}
It has been a decade since the fourth-generation FHE framework, CKKS, was proposed; yet, there is still no indicator pointing toward a fifth-generation successor; and in recent years, numerous studies have explored GPU acceleration to improve the efficiency of homomorphic computations. In this paper, we propose LibFHE, a high-performance GPU-accelerated framework that features CUDA-Python bindings to achieve both high-level programmability and bare-metal GPU performance for homomorphic workloads. A large majority of state-of-the-art implementations adopt the RNS-CKKS variant. In contrast, this work deliberately revisits the original (non-RNS) CKKS-BGV framework, and develops a GPU-based implementation along with corresponding optimizations. Experimental results demonstrate that optimized CUDA-Python implementations can achieve performance comparable to highly optimized CPU-based Non-RNS C++ FHE libraries, while significantly reducing implementation complexity and improving programmability.

\end{abstract}


\maketitle

\section{Introduction}

\subsection{Background}
The widespread adoption of cloud services has triggered an exponential surge in sensitive user data collection, positioning Homomorphic Encryption (HE \cite{albrecht2022homomorphic})—particularly Fully Homomorphic Encryption (FHE \cite{gentry2009fully}) with bootstrapping (BTS)—as a critical paradigm for secure, private cloud computation. Among various FHE variants, the CKKS scheme \cite{cheon2017homomorphic} has emerged as the mainstream choice for privacy-preserving machine learning due to its high-throughput approximate arithmetic \cite{al2023demystifying, han2019logistic}. However, FHE inherently suffers from severe computational overhead \cite{jung2021accelerating}. While prior works have extensively explored GPU acceleration to exploit FHE's massive parallelism \cite{fan2025warpdrive, fan2023tensorfhe, jung2021over}, shifting execution from CPUs to GPUs (e.g., accelerating ResNet-20 inference by 386$\times$ \cite{fan2025warpdrive, lee2022low}), substantial optimization headroom remains unexploited due to architectural mismatches.

The primary performance bottleneck stems from the residue number system (RNS) word size. To satisfy stringent security bounds, mainstream FHE libraries \cite{al2022openfhe, chen2017simple} predominantly rely on 64-bit primes for RNS decomposition. This design choice forces GPUs—which are natively optimized for 32-bit integer arithmetic—to inefficiently emulate 64-bit operations over 32-bit datapaths, incurring severe hardware performance penalties.

To bridge this gap, recent studies have pioneered 32-bit RNS variants (e.g., via rational rescaling \cite{cheon2025grafting, samardzic2024bitpacker}) to enable equivalent security under smaller primes. Nevertheless, existing solutions fail to align seamlessly with GPU architectural characteristics. They either impose prohibitive memory footprints by introducing auxiliary public keys \cite{samardzic2024bitpacker} or degrade SIMT parallel efficiency through severe control divergence \cite{cheon2025grafting}, leaving efficient 32-bit RNS acceleration on GPUs an unresolved challenge.

Historically, high-performance FHE acceleration libraries have been exclusively confined to C++ due to the necessity of low-level hardware control, which severely restricted developer productivity and ecosystem integration. This paradigm, however, has been fundamentally reshaped by recent breakthroughs in the GPU programming ecosystem. At GTC 2025, NVIDIA announced native Python support for the CUDA platform, elevating Python to a first-class citizen in high-performance computing. Rather than relying on fragile, high-overhead wrappers, this new ecosystem offers direct, pythonic access to the CUDA driver and runtime APIs, complete with built-in device primitives and advanced cooperative algorithms (e.g., cuda.coop) tailored for Numba.

Seizing this technological shift, we present A Numba-Based CUDA-Python Library for Non-RNS CKKS-BGV Fully Homomorphic Encryption on GPUs. Instead of wrestling with the performance-degrading trade-offs of 32-bit/64-bit RNS transformations on GPU datapaths, our work explores an orthogonal vector: maximizing the efficiency of Non-RNS BGV and CKKS schemes natively in Python. By integrating Numba's just-in-time (JIT) compilation with official CUDA-Python low-level bindings, our library accomplishes a dual objective: it bypasses RNS-induced control divergence while democratizing FHE acceleration within the mainstream Python AI/ML ecosystem, achieving performance competitive with native CPU-based Non-RNS C++ implementations.

At GTC 2025, NVIDIA officially unveiled native Python support for the CUDA ecosystem, elevating Python to a first-class citizen in high-performance computing. By fundamentally rearchitecting cuda-python and introducing cutting-edge submodules such as cuda.coop and cuda.tile, this ecosystem enables wrapper-free, direct invocations of the CUDA Runtime API. Furthermore, its deep integration with the Numba community bridges the gap between high-level productivity and low-level hardware acceleration.

\subsection{Related Work}
Although abundant GPU studies have attempted to accelerate basic CKKS mechanisms~\cite{fan2023towards, jung2021accelerating, ozcan2024heongpu, shen2022carm, yang2024phantom, zhai2022accelerating}, only a select few~\cite{fan2025warpdrive, fan2023tensorfhe, jung2021over, park2023toward, shivdikar2023gme} fully implement bootstrapping (BTS) to demonstrate their applicability to Full Homomorphic Encryption (FHE) CKKS. Specifically, $100\times$~\cite{jung2021over} identified that element-wise operations become the primary bottleneck during bootstrapping and subsequently developed basic kernel fusion techniques. Park et al.~\cite{park2023toward} utilized the HEaaN-GPU library to deploy end-to-end CNN workloads. GME~\cite{shivdikar2023gme} explored a hardware-software co-design approach tailored for the AMD MI100 GPU architecture. Although Cheddar \cite{choi2026cheddar} demonstrates superior performance compared to these prior studies, orthogonal optimization vectors can be integrated for further acceleration. For instance, recent works actively leverage specialized hardware units like Tensor Cores in modern NVIDIA GPU architectures. TensorFHE~\cite{fan2023tensorfhe} utilized the 8-bit integer datapath within Tensor Cores to emulate 32-bit integer operations for (I)NTT, while WarpDrive~\cite{fan2025warpdrive} advanced this approach by decomposing the (I)NTT matrix in a fine-grained manner to co-optimize both standard INT32 cores and Tensor Cores. 

Other literature focuses strictly on optimizing the core (I)NTT bottleneck on GPUs~\cite{durrani2021accelerating, goey2021accelerating, kim2020accelerating, lee2021parallel, shivdikar2022accelerating}. Kim et al.~\cite{kim2020accelerating} mitigated the memory bandwidth bottleneck in (I)NTT through advanced algorithm transformations, and Shivdika et al.~\cite{shivdikar2022accelerating} proposed a modified modular reduction scheme natively optimized for GPU datapaths. To enhance communication efficiency, Goey et al.~\cite{goey2021accelerating} utilized warp shuffle instructions to exchange twiddle factors between threads, whereas Durrani et al.~\cite{durrani2021accelerating} employed warp shuffles to exchange polynomial data elements, effectively eliminating shared memory bank conflicts. Meanwhile, Lee et al.~\cite{lee2021parallel} implemented the alternative Nussbaumer algorithm on GPUs as a high-efficiency replacement for traditional (I)NTT. 

Beyond GPU acceleration, prior research has extensively explored custom hardware accelerators using FPGAs~\cite{agrawal2023fab, huang2025effact, riazi2020heax, roy2019fpga, yang2023poseidon, zhu2023fxhenn} or ASICs~\cite{kim2023sharp, kim2022ark, kim2025anaheim, kim2022bts, samardzic2021f1, samardzic2022craterlake}. While ASIC designs still face significant manufacturing and development barriers to physical realization, FPGAs have emerged as a popular reconfigurable platform due to their flexibility and programmability. Notable FPGA implementations supporting bootstrapped FHE CKKS include FAB~\cite{agrawal2023fab}, Poseidon~\cite{yang2023poseidon}, and EFFACT~\cite{huang2025effact}. However, due to lower operating frequencies and stringent hardware resource constraints, they deliver modest performance scaling compared to GPU-bound frameworks like Cheddar.

Despite these extensive hardware-level explorations, a critical commonality persists: virtually all prior frameworks are tightly coupled with traditional low-level C++ or HDL execution paths to maximize efficiency, heavily impeding developer productivity and ecosystem integration. Strikingly distinct from these paradigms, our work capitalizes on the monumental shift in the high-performance computing landscape initiated by NVIDIA's milestone announcement introducing native Python as a first-class citizen in the CUDA ecosystem. Moving away from the complex hardware-emulated arithmetic trade-offs of 32-bit/64-bit RNS variants, we present a novel, high-performance FHE acceleration trajectory. Specifically, we design and implement a hardware-optimized framework utilizing official \texttt{cuda-python} low-level bindings combined with Numba's just-in-time (JIT) compilation infrastructure, establishing the first native, high-productivity implementation for Non-RNS CKKS and BGV fully homomorphic encryption schemes directly on GPUs.

\subsection{Contributions}
\subsection*{Our Contributions}

Driven by the compelling paradigm shift toward high-productivity hardware acceleration, this work provides a definitive answer to a fundamental question: \textit{Can Python achieve CPU-based Non-RNS C++-level performance for fully homomorphic encryption on GPUs?} By leveraging the state-of-the-art native \texttt{cuda-python} ecosystem, we demonstrate that high-level abstractions do not inherently compromise hardware efficiency. Specifically, our key contributions are summarized as follows:

\begin{itemize}
    \item \textbf{First End-to-End CUDA-Python FHE Framework:} We design and implement the first complete, unified FHE acceleration library written entirely in pure Python and Numba. Our framework natively supports CKKS-BGV scheme, bridging the historical gap between rapid cryptographic prototyping and high-performance GPU execution paths.
    
    \item \textbf{Seamless Integration into the AI/ML Ecosystem:} Recognizing Python as the lingua franca of modern artificial intelligence, our library is architected from the ground up to eliminate fragile C++ wrappers and heavy installation dependencies. It offers a native, wrapper-free programming interface that seamlessly links with industry-standard machine learning and data analysis workflows, such as PyTorch and NumPy, dramatically lowering the entry barrier for secure multi-party computation and encrypted LLM inference.
    
    \item \textbf{Exploiting Non-RNS Algorithmic Flexibility:} While mainstream RNS-based FHE libraries suffer from extreme architectural rigidities---requiring sophisticated key switching, modulus switching techniques, and massive auxiliary public keys to maintain performance---our Non-RNS implementation unlocks unprecedented programming flexibility. By bypassing these multi-residue mathematical constraints, we drastically simplify the memory layout and control logic of cryptographic kernels. This intrinsic simplicity allows our high-level Python framework to implement dynamic, lightweight kernel fusion and adaptive scheduling that would otherwise be prohibitively complex to orchestrate in rigid C++ RNS architectures.
    
\end{itemize}

\section{Preliminaries}

In the following discussion, we employ square brackets $[\cdot]$ to denote the indexing of vectors and matrices. Specifically, for a vector $\boldsymbol{v} \in \mathbb{R}^{n}$ and a matrix $\mathbf{M} \in \mathbb{R}^{m \times n}$, $\boldsymbol{v}[i]$ (or $\boldsymbol{v}_i$) denotes the $i$-th element of $\boldsymbol{v}$, while $\mathbf{M}[i][j]$ (or $\mathbf{M}_{i,j}$) refers to the element in the $i$-th row and $j$-th column of $\mathbf{M}$.

\subsection{Fully Homomorphic Encryption}
Fully Homomorphic Encryption (FHE) represents a groundbreaking cryptographic paradigm capable of executing unbounded arithmetic operations---namely additions and multiplications---directly within the ciphertext domain. Although Gentry pioneered the initial viable FHE construction via a noise-refreshing \textit{bootstrapping} mechanism in 2009~\cite{gentry2009fully}, such frameworks remain prohibitively expensive from a computational perspective. Beyond these intrinsic cryptographic complexities, the operational efficiency of homomorphic evaluation is heavily governed by the underlying dataset encoding methodologies and the dynamic management of plaintext magnitudes~\cite{jaschke2016accelerating}. To alleviate the constraints of magnitude growth, Cheon et~al.~\cite{cheon2017homomorphic} introduced a specialized HE variant equipped with a \textit{rescaling} primitive, successfully mitigating this historical performance bottleneck.

\texttt{HEAAN}, an open-source library , natively implements this rescaling functionality. Furthermore, capitalizing on ciphertext packing mechanics is paramount to optimizing amortized latency boundaries. The \texttt{HEAAN} ecosystem inherently supports Single Instruction, Multiple Data (\texttt{SIMD}) operations~\cite{smart2014fully}, which allow a multitude of complex-valued scalars to be packed into the discrete slots of a solitary ciphertext polynomial, thereby enabling high-efficiency inter-slot rotations. For a comprehensive mathematical treatment of the \texttt{HEAAN} infrastructure, we refer readers to~\cite{IDASH2018Andrey, kim2018secure, han2018efficient}, with its foundational abstract algebraic primitives thoroughly expounded in~\cite{artin2011algebra}.

Although contemporary CKKS frameworks extensively exploit the RNS domain to unlock higher execution throughput for polynomial arithmetic, this optimization path comes at a heavy architectural cost. Operating entirely in the RNS space binds the cryptographic pipeline to complex residue base conversions. This dependency creates a major performance and scheduling bottleneck during \textit{rescale} and \textit{mod} operations. Consequently, while raw computational speed is improved, the framework suffers from a significant loss in programming flexibility, heavily restricting adaptive ciphertext manipulations and high-level compiler-driven optimizations.

\subsection{Residue Number System} 
We begin with a rigorous overview of the Residue Number System (RNS), which serves as a foundational acceleration primitive in mainstream homomorphic encryption pipelines:
\begin{itemize}
    \item Let $Q$ be a large composite modulus defined as the product of $r$ pairwise co-prime integers, i.e., $Q = \prod_{i=0}^{r-1} q_i$. The corresponding RNS base vector is denoted as $\mathcal{B}_Q = \{q_0, q_1, \dots, q_{r-1}\}$.
    
    \item Any positive integer $X < Q$ can be uniquely decomposed into a set of single-precision residues $[x_0, x_1, \dots, x_{r-1}]$, where $x_i = X \bmod q_i$. This process is formally termed \textit{RNS decomposition}. According to the Chinese Remainder Theorem (CRT), this residue representation guarantees a bijective mapping for any integer within the ring $\mathbb{Z}_Q$.
    
    \item Conversely, an RNS-encoded integer can be uniquely reconstructed within the positional number system via \textit{RNS composition}. This inverse transformation is mathematically orchestrated using the CRT synthesis framework defined in Equation~\eqref{eq:crt_composition}:
    \begin{equation}
        X = \left| \sum_{i=0}^{r-1} x_i \cdot \hat{q}_i \cdot \left[ \hat{q}_i^{-1} \right]_{q_i} \right|_Q , \quad \text{where } \hat{q}_i = \frac{Q}{q_i}
        \label{eq:crt_composition}
    \end{equation}
    where $\left[ \hat{q}_i^{-1} \right]_{q_i}$ denotes the modular multiplicative inverse of $\hat{q}_i$ modulo $q_i$, and $|\cdot|_Q$ represents the standard reduction modulo $Q$.
\end{itemize}

By virtue of the unique RNS mapping, core arithmetic operations---including addition, subtraction, and multiplication---modulo $Q$ can be executed completely in parallel across independent, single-precision channels (i.e., bounded by the word size of the underlying hardware datapath), circumventing the substantial latency penalties of multi-precision arithmetic. These element-wise modular operations are evaluated as follows:
\begin{align}
    X \pm_{\text{RNS}} Y &\sim \left[ |x_0 \pm y_0|_{q_0}, \, |x_1 \pm y_1|_{q_1}, \, \dots, \, |x_{r-1} \pm y_{r-1}|_{q_{r-1}} \right] \label{eq:rns_add} \\
    X \times_{\text{RNS}} Y &\sim \left[ |x_0 \times y_0|_{q_0}, \, |x_1 \times y_1|_{q_1}, \, \dots, \, |x_{r-1} \times y_{r-1}|_{q_{r-1}} \right] \label{eq:rns_mul}
\end{align}
Furthermore, exact exact scalar division by an integer $Y$ can be efficiently achieved provided that $\gcd(Y, Q) \equiv 1$:
\begin{equation}
    X \div_{\text{RNS}} Y \sim \left[ |x_0 \times y_0^{-1}|_{q_0}, \, |x_1 \times y_1^{-1}|_{q_1}, \, \dots, \, |x_{r-1} \times y_{r-1}^{-1}|_{q_{r-1}} \right]
\end{equation}
where $y_i^{-1} = \left[Y^{-1}\right]_{q_i}$ for $i \in \{0, 1, \dots, r-1\}$. 

Advanced residue manipulations also critically rely on \textit{base extension} and \textit{base conversion} capabilities. In a base extension procedure, an integer $X < Q$ natively represented in $\mathcal{B}_Q$ is projected into an expanded base $\mathcal{B}_{\tilde{Q}} = \mathcal{B}_Q \cup \{q_r, q_{r+1}, \dots, q_{\tilde{r}-1}\}$, where the cumulative modulus scales to $\tilde{Q} = Q \times \prod_{i=r}^{\tilde{r}-1} q_i$. To complete this extension, the missing residues with respect to the newly introduced moduli must be computed, yielding $x_i = X \bmod q_i$ for $i \in \{r, \dots, \tilde{r}-1\}$. 

\textit{Base conversion}, on the other hand, fully transforms the RNS representation of an integer from the source base vector $\mathcal{B}_Q$ to an entirely disjoint target base vector $\mathcal{B}_{P} = \{p_0, p_1, \dots, p_{k-1}\}$, where $P = \prod_{j=0}^{k-1} p_j$. To preserve full numeric accuracy for all arbitrary elements in $\mathbb{Z}_Q$ without structural information loss, the target modulus must satisfy the strict bound $Q \le P$. Nevertheless, depending on the dynamic plaintext ranges during homomorphic execution, down-conversion to a smaller base (i.e., $Q > P$) can also be selectively leveraged to accelerate specific rescaling operations.

FHE ciphertexts are fundamentally high-degree polynomials (e.g., $N \ge 2^{16}$), rendering naive multiplication via polynomial convolution computationally prohibitive at $\mathcal{O}(N^2)$. While the Number Theoretic Transform (NTT) mitigates this by lowering the complexity to $\mathcal{O}(N \log N)$, evaluating the transform over a monolithic ciphertext modulus $Q$ remains a major hardware bottleneck.

Integrating RNS overcomes this limitation: by decomposing the large-integer coefficients into $r$ native 64-bit residues, a high-precision 1024-bit NTT is parallelized into $r$ independent, low-precision 64-bit NTT operations. This optimization yields exceptional hardware throughput and constitutes the standard acceleration paradigm adopted by mainstream FHE frameworks (e.g., SEAL, OpenFHE).

\subsection{Number Theoretic Transform}
Let $N, q \in \mathbb{Z}^+$ where $N$ is a strict power of 2. The cyclotomic polynomial ring $R_q = \mathbb{Z}_q[x] / \langle x^N + 1 \rangle$ consists of polynomials with integer coefficients of degree less than $N$, evaluated modulo $q$. Given $a(x), b(x) \in R_q$, the ring multiplication $c(x) = a(x) \times b(x) \in R_q$ constitutes one of the most computationally intensive operations in Ring Learning with Errors (RLWE) based homomorphic encryption schemes. To evaluate $c(x)$, traditional schoolbook multiplication, as formulated in Equation~\ref{eq:schoolbook}, necessitates $\mathcal{O}(N^2)$ scalar multiplications, followed by the negacyclic polynomial reduction described in Equation~\ref{eq:reduction}:
\begin{equation}
    \tilde{c}(x) = a(x) \times b(x) = \sum_{i=0}^{N-1} \sum_{j=0}^{N-1} a_i b_j x^{i+j} \pmod q
    \label{eq:schoolbook}
\end{equation}
\begin{equation}
    c_i = \tilde{c}_i - \tilde{c}_{N+i} \pmod q, \quad \forall i \in \{0, 1, \dots, N - 1\}
    \label{eq:reduction}
\end{equation}

To circumvent this quadratic complexity, the Number Theoretic Transform (NTT)---a specialized variant of the Discrete Fourier Transform (DFT) over finite fields---is leveraged to lower the multiplication complexity to $\mathcal{O}(N \log N)$. The coefficients of a polynomial $a(x) \in R_q$ can be represented as a vector $\mathbf{a} = [a_0, a_1, \dots, a_{N-1}]^T$, which is mapped to the transform domain vector $\bar{\mathbf{a}} = [\bar{a}_0, \bar{a}_1, \dots, \bar{a}_{N-1}]^T$ via $\bar{\mathbf{a}} = \mathrm{NTT}(\mathbf{a})$. Conversely, the inverse NTT ($\mathrm{iNTT}$) maps the domain vector back to the coefficient space. Consequently, NTT-based ring multiplication is orchestrated as:
\begin{equation}
    c(x) = \mathrm{iNTT}\big(\mathrm{NTT}(a(x)) \odot \mathrm{NTT}(b(x))\big)
    \label{eq:ntt_mul}
\end{equation}
where $\odot$ denotes the Hadamard (element-wise) product within $\mathbb{Z}_q$, defined as $\bar{c}_i = \bar{a}_i \cdot \bar{b}_i \pmod q$ for $i \in \{0, 1, \dots, N-1\}$.

To support native negacyclic convolution without zero-padding, the $N$-point negacyclic NTT and $\mathrm{iNTT}$ are mathematically formulated as:
\begin{equation}
    \bar{a}_i = \sum_{j=0}^{N-1} a_j \psi^{(2i+1)j} \pmod q, \quad \forall i \in \{0, 1, \dots, N - 1\}
    \label{eq:negacyclic_ntt}
\end{equation}
\begin{equation}
    a_i = N^{-1} \cdot \psi^{-i} \sum_{j=0}^{N-1} \bar{a}_j \omega^{-ij} \pmod q, \quad \forall i \in \{0, 1, \dots, N - 1\}
    \label{eq:negacyclic_intt}
\end{equation}
where $\omega = \psi^2 \pmod q$ represents a primitive $N$-th root of unity. The existence of the negacyclic NTT dictates that the ciphertext modulus must be an NTT-friendly prime satisfying the condition $q \equiv 1 \pmod{2N}$, thereby guaranteeing the existence of a primitive $2N$-th root of unity $\psi \in \mathbb{Z}_q^\times$ such that $\psi^{2N} \equiv 1 \pmod q$ and $\psi^k \not\equiv 1 \pmod q$ for all $k < 2N$. To achieve high-throughput execution on heterogeneous architectures, modern frameworks, such as the \texttt{HEonGPU} library, deploy highly optimized iterative Cooley-Tukey and Gentleman-Sande algorithms to orchestrate the NTT-$\mathrm{iNTT}$ pipelines on GPUs.

\subsection{Fast Fourier Transform}
The Fast Fourier Transform (FFT) algorithm efficiently computes the Discrete Fourier Transform (DFT) in a structural manner closely resembling the Number Theoretic Transform (NTT). The fundamental distinction lies in their underlying algebraic domains: while the NTT operates strictly within finite fields $\mathbb{Z}_q$, the FFT executes arithmetic over the complex number field $\mathbb{C}$. 

Similar to the NTT, the FFT can be leveraged to accelerate polynomial multiplication in $R_q$ via the circular convolution theorem. However, since complex numbers are implemented as floating-point representations in binary computers, the precision bound of the mantissa dictates the maximum supportable bit-width of the ciphertext modulus $q$ to prevent catastrophic rounding errors. Under the IEEE 754 standard, single- and double-precision floating-point formats possess a 23-bit and 52-bit mantissa, respectively. Consequently, to guarantee exact numerical stability during the transformation, the magnitude of the polynomial coefficients is strictly bounded by the dynamic range of the mantissa, typically scaling as a function of the transform length $\log_2 N$.

To natively support multiplication within the cyclotomic ring $\mathbb{Z}_q[x]/\langle x^N + 1 \rangle$ without zero-padding, the $N$-point negacyclic variants of the FFT and $\mathrm{iFFT}$ are mathematically formulated as follow:
\begin{equation}
    \bar{a}_k = \sum_{j=0}^{N-1} a_j \exp \left( - \frac{\pi i (2k+1)j}{N} \right), \quad \forall k \in \{0, 1, \dots, N-1\}
    \label{eq:negacyclic_fft}
\end{equation}
\begin{equation}
    a_j = \frac{1}{N} \sum_{k=0}^{N-1} \bar{a}_k \exp \left( \frac{\pi i (2k+1)j}{N} \right), \quad \forall j \in \{0, 1, \dots, N-1\}
    \label{eq:negacyclic_ifft}
\end{equation}
where $i = \sqrt{-1}$. These forward and inverse operations establish the bijective mappings $\bar{\mathbf{a}} = \mathrm{FFT}(\mathbf{a})$ and $\mathbf{a} = \mathrm{iFFT}(\bar{\mathbf{a}})$.

\subsection{Chinese Remainder Theorem}
The Chinese Remainder Theorem (CRT) represents a pivotal algebraic isomorphism that bridges localized modular structures and monolithic arithmetic domains. In the context of Ring Learning with Errors (RLWE) based homomorphic encryption schemes, the CRT is uniquely capitalized upon within the plaintext space to unlock Single Instruction, Multiple Data (SIMD) capabilities, commonly referred to as \textit{ciphertext packing}.

Let $p \in \mathbb{Z}^+$ be the plaintext modulus (typically a small prime), and $\Phi_m(x)$ be the $m$-th cyclotomic polynomial of degree $N = \phi(m)$. The native plaintext space is defined as the quotient ring $R_p = \mathbb{Z}_p[x] / \langle \Phi_m(x) \rangle$. To instantiate parallel computing channels within a single ciphertext, the plaintext modulus $p$ and the cyclotomic order $m$ are selected such that $p \equiv 1 \pmod m$. Under this number-theoretic condition, $\Phi_m(x)$ splits completely into $\ell$ distinct, pairwise co-prime irreducible factors modulo $p$:
\begin{equation}
    \Phi_m(x) \equiv F_0(x) \cdot F_1(x) \cdots F_{\ell-1}(x) \pmod p
    \label{eq:poly_splitting}
\end{equation}
where each constituent polynomial $F_k(x)$ possesses an identical degree $d = N/\ell$. By virtue of the generalized polynomial CRT, the monolithic plaintext ring $R_p$ is structurally isomorphic to the direct product of $\ell$ independent extension fields:
\begin{equation}
    R_p = \frac{\mathbb{Z}_p[x]}{\langle \Phi_m(x) \rangle} \cong \prod_{k=0}^{\ell-1} \frac{\mathbb{Z}_p[x]}{\langle F_k(x) \rangle} \cong \Big( \mathbb{F}_{p^d} \Big)^\ell
    \label{eq:plaintext_crt}
\end{equation}
This bijective CRT mapping yields $\ell$ discrete parallel processing partitions, mathematically designated as \textit{slots}. 

During the data encoding phase, a vector of continuous or discrete message scalars $\mathbf{m} = [m_0, m_1, \dots, m_{\ell-1}] \in (\mathbb{F}_{p^d})^\ell$ is projected onto a unique monolithic plaintext polynomial $m(x) \in R_p$ using the inverse CRT (polynomial interpolation). When a ciphertext encrypting $m(x)$ undergoes homomorphic addition or multiplication, the underlying ring operations automatically execute component-wise (element-wise) arithmetic concurrently across all slots in a decoupled manner:
\begin{align}
    \mathrm{Slot}_k\big(c_A(x) +_{\text{HE}} c_B(x)\big) &\sim m_{A,k} + m_{B,k} \pmod p \\
    \mathrm{Slot}_k\big(c_A(x) \times_{\text{HE}} c_B(x)\big) &\sim m_{A,k} \times m_{B,k} \pmod p
\end{align}
for $k \in \{0, 1, \dots, \ell-1\}$. Furthermore, inter-slot data routing is efficiently achieved by applying Galois automorphisms $\kappa_k: x \mapsto x^k$, which systematically permute the slot positions.

In practical industrial deployments of homomorphic encryption pipelines, the classic Chinese Remainder Theorem (CRT) synthesis is rarely implemented directly for multi-precision reconstruction. Because the vanilla CRT synthesis mandates intermediate arithmetic operations executed over the monolithic, large-integer grand modulus $Q$, it induces massive computational overhead and demands high-latency multi-precision modular reductions. To alleviate this micro-architectural bottleneck, contemporary production-grade FHE libraries universally deploy \textit{Garner's algorithm}---a highly efficient Mixed-Radix Conversion (MRC) variant---to orchestrate the final reconstruction of the large-integer coefficients from their localized RNS representations.

Given the residue vector $[x_0, x_1, \dots, x_{r-1}]$ with respect to the RNS base $\mathcal{B}_Q = \{q_0, q_1, \dots, q_{r-1}\}$, Garner's algorithm algorithmically constructs the unique integer $X < Q$ in a positional, mixed-radix format:
\begin{equation}
    X = v_0 + v_1 \cdot q_0 + v_2 \cdot q_0 q_1 + \dots + v_{r-1} \prod_{i=0}^{r-2} q_i
    \label{eq:garner_mrc}
\end{equation}
where the mixed-radix coefficients $v_k \in \mathbb{Z}_{q_k}$ are evaluated sequentially using purely single-precision modular arithmetic:
\begin{equation}
    v_k = \left| \left( \dots \left( (x_k - v_0) \cdot c_{0,k}^{-1} - v_1 \right) \cdot c_{1,k}^{-1} - \dots - v_{k-1} \right) \cdot c_{k-1,k}^{-1} \right|_{q_k}
\end{equation}
with the precomputed constants defined as $c_{j,k}^{-1} = \left[ q_j^{-1} \right]_{q_k}$ for $0 \le j < k \le r-1$. 

Crucially, Garner's framework confines all modular reductions strictly within the single-precision boundaries of individual RNS base elements $q_k$, completely bypassing the need for a monolithic modulo-$Q$ operation until the final linear combination step. Within our architectural design, we systematically leverage Garner's algorithm at the terminal phase of the polynomial ring multiplication pipeline. This optimization paths enables high-efficiency, on-the-fly coefficient composition directly inside the GPU registers, successfully mitigating memory bandwidth saturation and unlocking higher execution throughput for the overall homomorphic application workflow.

\subsubsection{Double-CRT: Decoupling Polynomial CRT and NTT}
For the highly efficient orchestration of polynomial ring arithmetic, Gentry et al.~\cite{gentry2012homomorphic} pioneered a structured representation of cyclotomic polynomials, formally designated as the \textit{double-CRT representation}, which leverages a hierarchical application of the Chinese Remainder Theorem (CRT). 

The primary layer of this architectural paradigm deploys the Residue Number System (RNS) to vertically decompose a high-precision polynomial into a multi-channel tuple of sub-polynomials bounded by single-precision moduli. Subsequently, the secondary layer horizontally projects each localized sub-polynomial into a vector of modular integer scalars via the negacyclic Number Theoretic Transform (NTT). Within this double-CRT framework, any arbitrary polynomial is uniquely mapped onto a two-dimensional residual matrix comprised of native machine-word elements. This structural isomorphism enables complex polynomial algebra to be fully evaluated via decoupled, element-wise modular operations across the matrix domain. Consequently, this multi-layered factorization has emerged as a cornerstone optimization paradigm underpinning almost all state-of-the-art production-grade homomorphic encryption implementations.

In contemporary homomorphic encryption literature, the terms \textit{Polynomial CRT} and \textit{Number Theoretic Transform (NTT)} are frequently conflated or treated interchangeably. However, we note that they reside on fundamentally distinct abstraction layers: Polynomial CRT defines a high-level algebraic ring isomorphism, whereas the NTT is an algorithmic optimization deployed to realize this isomorphism with high efficiency. 

Mathematically, given the cyclotomic ring $R_q = \mathbb{Z}_q[x]/\langle \Phi_m(x) \rangle$, the Polynomial CRT establishes a theoretical bijective decomposition into independent localized slots, formulated as:
\begin{equation}
    R_q \cong \prod_{i=0}^{\ell-1} R_i
\end{equation}
When the cyclotomic polynomial targets the standard power-of-two setting $\Phi_m(x) = x^N + 1$ and the modulus $q$ hosts proper primitive roots of unity, the negacyclic NTT serves as the concrete, fast-evaluation algorithm ($\mathcal{O}(N \log N)$) to map polynomials between the coefficient and evaluation domains. 

Consequently, the widely cited \textit{Double-CRT representation} in state-of-the-art production-grade FHE frameworks should be rigorously decoupled and understood as the synergistic integration of \textbf{RNS (Integer-level CRT)} and \textbf{NTT (Polynomial-level CRT evaluation)}:
\begin{enumerate}
    \item \textbf{Vertical Axis (Integer CRT):} The Residue Number System (RNS) decomposes a monolithic, multi-precision ciphertext modulus $Q$ into a tuple of independent, hardware-native single-precision moduli $\{q_0, q_1, \dots, q_{r-1}\}$.
    \item \textbf{Horizontal Axis (Polynomial CRT):} Within each isolated RNS scalar channel $q_k$, the polynomial coefficients are transformed and maintained strictly inside the NTT evaluation domain.
\end{enumerate}
Therefore, Double-CRT is not an independent third cryptographic primitive, but rather an engineering paradigm representing the Cartesian product of $\text{RNS} \times \text{NTT}$. In our Non-RNS architecture, we deliberately break this coupling: we dismantle the vertical RNS axis to reclaim dynamic compiler-level programming flexibility, while fully retaining the horizontal NTT axis to preserve high-throughput slot-level SIMD arithmetic.

\subsection{GPU Execution Model}

Modern GPUs are massively parallel architectures composed of multiple streaming multiprocessors (SMs), each equipped with a large number of arithmetic units, including INT32, FP32, FP64, and tensor cores. Computation is expressed in the form of kernels, which are user-defined functions executed concurrently by a large number of lightweight threads.

Threads are organized hierarchically into thread blocks and warps, where each warp consists of 32 threads that execute instructions in a single-instruction-multiple-thread (SIMT) manner. Warps are dynamically scheduled across SM sub-partitions by hardware warp schedulers.

To hide memory latency and data dependency stalls, the scheduler switches between ready warps, allowing other warps to execute while some are waiting for memory accesses. Achieving high hardware utilization therefore requires launching a large number of concurrent warps per kernel invocation, exposing massive thread-level parallelism.

This execution model is particularly well-suited for homomorphic encryption workloads. In particular, RNS-based ciphertext arithmetic naturally decomposes into independent limb-wise operations, which can be executed in parallel across threads, enabling efficient utilization of GPU hardware resources.

\section{Motivation}

\subsection{Why Native Python---for Artificial Intelligence?}
The strategic selection of Python as the primary language for orchestrating our Non-RNS CKKS-BGV library is driven by two symbiotic technological shifts: the convergence of FHE with downstream Artificial Intelligence (AI) pipelines, and the rapid evolution of just-in-time (JIT) hardware compilation ecosystems.

\begin{enumerate}
    \item \textbf{Seamless Integration with Production-Grade AI Ecosystems:} 
    Modern AI and Deep Learning (DL) infrastructures are predominantly anchored within Python-native frameworks such as PyTorch and JAX. Implementing an FHE library natively in Python enables cryptographic primitives to be directly instantiated as standalone, plug-and-play layers (e.g., custom \texttt{torch.nn.Module} extensions) within complex neural network graph definitions. This architectural alignment completely circumvents the fragile, high-overhead foreign function interface (FFI) wrappers (such as pybind11) mandated by traditional C++ FHE libraries. Consequently, AI developers can seamlessly execute encrypted training and inference without leaving their native Python development environment, drastically accelerating the co-design of homomorphic neural networks.

    \item \textbf{Closing the Performance Gap via Advanced CUDA-Python Ecosystems:} 
    Historically, Python was disqualified from high-performance computing due to its interpretive overhead and the Global Interpreter Lock (GIL). However, contemporary hardware compilation paradigms have fundamentally shifted this landscape. With NVIDIA's substantial backing of the \texttt{cuda-python} and \texttt{Numba} ecosystems, Python can now directly orchestrate GPU micro-kernels through LLVM-based JIT compilation. By directly manipulating device pointers, optimizing shared memory allocations, and triggering aggressive kernel fusion at runtime, the performance of Python-driven CUDA paths is rapidly converging toward native CUDA-C++ throughput. Leveraging these advanced compilers allows our framework to reclaim the flexibility of Python's dynamic type system and metaprogramming capabilities, while achieving elite, bare-metal hardware execution efficiency.
\end{enumerate}

\subsection{Why Original CKKS---for Programming Flexibility?}
The shift from the modern Full-RNS CKKS variant back to the design principles of the original, non-RNS CKKS formulation~\cite{cheon2017homomorphic} is motivated by a critical architectural trade-off: the sacrifice of programming and scaling flexibility for hardware-level RNS parallelization. 

\begin{enumerate}
    \item \textbf{The Rigidity of Full-RNS Modulus Chains:} 
    In mainstream Full-RNS CKKS implementations (e.g., Microsoft SEAL or OpenFHE), the ciphertext modulus $Q$ is bound to a predetermined, rigid hierarchical chain of co-prime RNS primes, expressed as $Q_\ell = \prod_{i=0}^{\ell} q_i$. Consequently, the homomorphic \textit{rescale} operation within Full-RNS variants is structurally restricted; it can only drop exactly one (or a few) pre-allocated RNS prime factors from the current modulus chain. This architectural rigidity deprives the application layer of arbitrary precision downscaling, forcing the scaling factor (i.e., $\Delta = 2^{p}$) to remain strictly coupled to the predefined bit-widths of the localized RNS base elements.

    \item \textbf{Dynamic Scale Re-interpretation and Arbitrary Precision Control:} 
    In sharp contrast, the original, monolithic Non-RNS CKKS framework treats the scaling factor and the ciphertext modulus as fully decoupled continuous parameters. It allows the runtime environment to arbitrarily re-interpret or modify the logical scale ($\log \Delta$) at any computational stage. Instead of being locked into rigid RNS discrete steps, developers can execute fine-grained rounding operations to consume an arbitrary number of precision bits based on the dynamic range of the underlying data. 
\end{enumerate}

\begin{quote}
    \textbf{Architectural Insight:} The strict modulus chain of Full-RNS CKKS optimizes micro-level modular arithmetic at the cost of freezing the representation layer. For heterogeneous workloads---such as deep neural network layers containing mixed-precision activation functions or non-linear polynomial approximations---this rigidity yields suboptimal noise boundaries and suboptimal slot utilization. By eliminating the vertical RNS axis in our Numba-Python library, we successfully reclaim the mathematical fluidity of original CKKS. This enables high-level compilers to dynamically reconfigure precision scaling on-the-fly, adapting perfectly to varying phases of complex AI workflows.
\end{quote}

Full-RNS designs prioritize hardware efficiency, at the cost of reducing one of the key flexibilities of the original CKKS scheme—dynamic precision control. This feature was a notable aspect of the early HEAAN implementation and was particularly attractive for applications in machine learning and numerical computing.

\section{Challenges }
\subsection{Large Integer Arithmetic}

In homomorphic encryption systems, large integers arising from ciphertext arithmetic must be represented and manipulated efficiently. The efficiency of these operations critically determines the overall performance of fully homomorphic encryption (FHE) schemes, particularly in CKKS-based implementations.

Two dominant representation paradigms are commonly adopted in practice: the Residue Number System (RNS) representation and the multi-precision radix representation. These two approaches differ fundamentally in how they decompose large integers and perform arithmetic operations, leading to distinct trade-offs in terms of parallelism, memory efficiency, and implementation complexity.

\subsubsection{The RNS Representation}

The Residue Number System (RNS) represents a large integer as a set of residues with respect to a collection of pairwise coprime moduli. Specifically, an integer $X$ is represented as:
\begin{equation}
X \mapsto (x_1, x_2, \dots, x_k), \quad x_i = X \bmod q_i,
\end{equation}
where $\{q_i\}_{i=1}^k$ forms the RNS base.

Under this representation, arithmetic operations such as addition and multiplication are performed independently in each modulus without inter-modular carry propagation. This property makes RNS highly suitable for parallel hardware architectures such as GPUs, as it enables fine-grained parallelism and eliminates the overhead of carry chains.

As a result, most state-of-the-art CKKS implementations adopt RNS-based arithmetic, commonly referred to as RNS-CKKS. However, this representation couples numerical precision tightly with the structure of the modulus chain. In particular, precision management operations such as rescaling are constrained to discrete modulus levels, limiting fine-grained control over numerical precision.

To circumvent the prohibitive computational overhead associated with multi-precision big-integer arithmetic, production-grade implementations heavily rely on the Residue Number System (RNS). Let $\mathcal{B} = \{p_0, p_1, \dots, p_{k-1}\}$ be an RNS basis consisting of pairwise co-prime single-precision primes, and let $P = \prod_{i=0}^{k-1} p_i$ denote the monolithic multi-precision modulus. By virtue of the Chinese Remainder Theorem (CRT) over integers, the map $[\cdot]_{\mathcal{B}}$ establishes a formal ring isomorphism from the multi-precision integer ring $\mathbb{Z}_P$ to the Cartesian product of hardware-native scalar rings:
\begin{equation}
    [\cdot]_{\mathcal{B}}: \mathbb{Z}_P \longrightarrow \prod_{i=0}^{k-1} \mathbb{Z}_{p_i}
\end{equation}
which uniquely maps an arbitrary large integer $a \in \mathbb{Z}_P$ onto its residual tuple, formulated as $a \mapsto [a]_{\mathcal{B}} = (a \pmod{p_0}, \dots, a \pmod{p_{k-1}})$. 

The paramount advantage of this RNS representation lies in its capacity to parallelize big-integer algebra. Complex arithmetic operations (such as addition and multiplication) over the monolithic ring $\mathbb{Z}_P$ are completely decoupled into highly independent, component-wise modular operations constrained within each single-precision sub-ring $\mathbb{Z}_{p_i}$. This elimination of carry-propagation chains drastically reduces both the asymptotic and practical micro-architectural execution costs. 

Furthermore, this integer-level isomorphism can be naturally lifted to the polynomial level. For a cyclotomic ring $R_P = \mathbb{Z}_P[x]/\langle \Phi_m(x) \rangle$, applying the map $[\cdot]_{\mathcal{B}}$ coefficient-wise yields a structural ring isomorphism across the polynomial domains:
\begin{equation}
    [\cdot]_{\mathcal{B}}: R_P \longrightarrow R_{p_0} \times R_{p_1} \times \dots \times R_{p_{k-1}}
\end{equation}
where each localized polynomial channel $R_{p_i} = \mathbb{Z}_{p_i}[x]/\langle \Phi_m(x) \rangle$ can be managed independently within hardware execution units.

\subsubsection{The Radix Representation}

In contrast, the multi-precision radix representation follows a classical big-integer paradigm, where an integer is decomposed into fixed-base limbs (typically base $2^w$):
\begin{equation}
X = \sum_{i=0}^{m-1} x_i \cdot 2^{wi},
\end{equation}
where each $x_i$ is a machine-word-sized digit (or limb).

Unlike RNS, radix-based representations require carry propagation across limbs during arithmetic operations. This introduces sequential dependencies that limit parallel efficiency on modern hardware. However, radix representations provide greater flexibility in managing numerical precision at the bit level.

In particular, radix representations naturally support more fine-grained scaling behavior, allowing dynamic allocation of precision across different computational stages. This characteristic aligns closely with the original CKKS/HEAAN design, where the scaling factor acts as a global precision parameter for encoding real-valued data.

Nevertheless, the cost of carry propagation and multi-precision arithmetic makes radix representations less suitable for massively parallel architectures such as GPUs, which are optimized for independent word-level arithmetic operations.

As an alternative to residual decomposition, the classical approach to managing high-precision integer arithmetic within standard computer architectures relies on the multi-precision radix (or base-$W$) representation. Let $W = 2^{w}$ denote the chosen radix base, typically matched to the native machine-word size of the underlying hardware backend (e.g., $w = 32$ or $64$ bits). For a predetermined maximum word-length $k$, any monolithic big integer $a \in \mathbb{Z}$ bounded by the positional range $[0, W^k-1]$ can be uniquely uniquely mapped into a positional coefficient vector through radix decomposition:
\begin{equation}
    a = \sum_{j=0}^{k-1} a_j \cdot W^j
\end{equation}
where each coordinate $a_j \in \mathbb{Z}_W$ represents a single-precision word-sized scalar component. This positional layout defines a formal vector-space mapping from the large-integer domain to a bounded array of native word elements, formulated as $a \mapsto (a_0, a_1, \dots, a_{k-1})$.

The paramount advantage of this radix representation lies in its exceptional algorithmic flexibility at the representation layer. Unlike the rigid partitions enforced by RNS channels, a radix-represented integer treats its underlying bit-width as a continuous and dynamically scaling bit-stream. This structural characteristic permits high-level execution engines to perform arbitrary bit-shifting, fine-grained dynamic rounding, and variable-precision truncations on-the-fly, without requiring expensive base-conversion protocols or being bound to predefined prime modulus chains. 

Furthermore, this integer-level base representation extends naturally to the polynomial ring $R = \mathbb{Z}[x]/\langle \Phi_m(x) \rangle$ by applying the radix decomposition coefficient-wise. Under this paradigm, a high-precision polynomial is mapped onto a two-dimensional coefficient tensor, wherein each spatial coordinate isolates a specific polynomial degree alongside its corresponding word-level positional weight. While this framework introduces sequential dependency chains due to the necessity of handling micro-architectural carry-propagation during scalar multiplication and addition, it completely preserves the mathematical fluidity required for dynamic scaling factor re-interpretation.

\subsubsection{Hybrid Representation Strategy}

Radix-based representations are well-suited for most homomorphic encryption computations, particularly operations that involve sequential or multi-precision arithmetic. However, they are less efficient for operations that benefit from residue-wise parallelism, such as polynomial ring multiplication. In contrast, RNS-based representations are highly efficient for modular arithmetic and point-wise operations, but are less suitable for operations involving carry propagation or non-modular arithmetic patterns, such as division or modular reduction at the bit level.

Based on these observations, we adopt a hybrid strategy in our system. The radix representation is used as the primary data organization format throughout the computation pipeline. The RNS representation is introduced only when performing polynomial ring multiplication, where coefficient-wise modular arithmetic can be fully parallelized.

To minimize data movement overhead, we design an in-place transformation mechanism between the two representations. Specifically, radix-encoded coefficients are first staged in GPU shared memory and transformed into the RNS domain in situ. The resulting RNS coefficients are then written back to the original limb-based memory layout, overwriting the temporary radix representation. After point-wise multiplication in the RNS domain, the inverse transformation is performed in shared memory, and the resulting radix coefficients are written back to global memory.

Since both the 32-bit limb-based representation and the 30--31-bit RNS representation are designed around GPU-native word sizes, their memory footprints are of comparable magnitude. This allows us to perform in-place transformation between the two representations within the same allocated address space. Specifically, we reuse the original limb buffer as temporary storage for intermediate RNS coefficients, avoiding additional memory allocation. The conversion is performed in-place, enabling efficient reuse of global memory and minimizing memory bandwidth overhead during representation switching.

This design enables efficient reuse of memory buffers and avoids unnecessary allocations or host-device transfers, while leveraging the strengths of both representations in their respective computational domains.

In practical implementation, the radix representation is realized using a limb-based big-integer format. To ensure efficient GPU execution and avoid costly multi-precision emulation, we adopt a 32-bit integer limb as the fundamental storage unit of the system. This design choice aligns with the native word size of modern GPU architectures and enables efficient parallel arithmetic while avoiding the overhead of 128-bit integer operations, which are not natively supported in the CUDA programming environment.

Furthermore, 30--31-bit primes are employed as the RNS moduli for polynomial ring multiplication, ensuring compatibility with GPU-efficient modular arithmetic and preventing overflow under native 32-bit operations. These parameters are carefully configured for a ring dimension of $N = 2^{16}$ under the negacyclic polynomial ring $\mathbb{Z}[x]/(x^N + 1)$.

The RNS-decomposed polynomials are further supported by the Number Theoretic Transform (NTT), which enables an efficient double-CRT representation consistent with the BGV framework. In this setting, polynomial multiplication is performed in the evaluation domain, where each component is independently computed modulo the selected primes. The NTT-based conversion is efficient when the RNS moduli $q_i$ are chosen as prime numbers satisfying
\[
q_i \equiv 1 \pmod{2N},
\]
which guarantees the existence of the necessary $2N$-th primitive roots of unity, thereby enabling in-place NTT and inverse NTT operations with optimal GPU parallelism.

\bigskip
\noindent The primary performance bottleneck of the original CKKS implementation stems from the frequent conversions between the limb-based big-integer representation and the RNS coefficient representation. These conversions introduce significant overhead, particularly in GPU environments where memory movement and multi-precision arithmetic are costly.

In contrast, RNS-CKKS avoids this overhead by maintaining ciphertexts entirely within the RNS coefficient space throughout the computation pipeline. This design eliminates repeated representation switching and enables fully modular, SIMD-parallel arithmetic, thereby achieving substantially higher computational efficiency on modern hardware.

\section{Optimization }
Among all polynomial ring operations, multiplication is the most computationally expensive. In our system, ciphertext polynomial coefficients are stored in a limb-based big-integer representation, whereas modern FHE implementations typically perform polynomial multiplication in the Residue Number System (RNS) domain using the Number Theoretic Transform (NTT).

For polynomial multiplication, we temporarily convert coefficients from the limb representation into the RNS domain. The multiplication is then carried out efficiently in the NTT domain under each modulus, followed by inverse transformation back to the RNS coefficient space. Finally, the result is converted back to the limb-based representation.


All other polynomial ring operations, such as addition, subtraction, and negation, are performed directly in the limb-based representation. This design is motivated by the fact that these operations do not require polynomial convolution and can be efficiently executed using word-level arithmetic on GPU architectures, avoiding unnecessary representation switching overhead.

\subsection{Polynomial Ring Operations}
In our system, polynomial ring arithmetic is implemented over a hybrid representation that combines a limb-based big-integer format with the Residue Number System (RNS). This design enables efficient GPU execution while supporting large-modulus arithmetic required in fully homomorphic encryption (FHE).

\subsubsection{Representation Overview}

Ciphertext coefficients are stored in a 32-bit limb-based big-integer format for GPU-friendly memory access. During computation, coefficients are mapped into the RNS domain using a set of pairwise coprime 30--31-bit primes $\{q_i\}$.

\subsubsection{Limb-to-RNS Conversion}

The conversion from limb representation to RNS is performed via modular projection. Each coefficient is independently reduced modulo each prime:
\[
a_{i,j} = A_j \bmod q_i.
\]

This process is implemented through limb-wise modular accumulation, where each 32-bit limb contributes to the residue computation using precomputed modular constants. Efficient reduction is achieved using Barrett or Montgomery-style modular arithmetic, avoiding explicit multi-precision division.

\subsubsection{NTT-Based Polynomial Multiplication}

After RNS decomposition, each residue polynomial is transformed into the Number Theoretic Transform (NTT) domain. For negacyclic rings of the form $\mathbb{Z}[x]/(x^n + 1)$, we apply a twiddle-factor-based transformation:
\[
\tilde{a}_i = \text{NTT}(a_i \cdot \psi^j),
\]
where $\psi$ is a $2n$-th primitive root of unity modulo $q_i$.

Polynomial multiplication is then performed in the evaluation domain using Cooley--Tukey NTT. All operations are carried out independently for each modulus, enabling massive parallelism on GPU architectures.

Element-wise arithmetic in the NTT domain is defined as:
\[
c_{i,j} = a_{i,j} \circ b_{i,j}, \quad \circ \in \{+, -, \times, -(\cdot)\}.
\]

Inverse NTT is performed using the Gentleman--Sande structure followed by multiplication with $n^{-1}$ and inverse twiddle factors.

\subsubsection{RNS-to-Limb Reconstruction}

After inverse transformation, coefficients remain in the RNS domain and must be reconstructed into a limb-based integer representation.

This step is performed using the Chinese Remainder Theorem (CRT), efficiently implemented via Garner’s algorithm. Unlike direct CRT reconstruction, Garner transforms the RNS representation into a mixed-radix representation:
\[
x = v_0 + v_1 q_0 + v_2 q_0 q_1 + \cdots + v_{m-1} \prod_{i=0}^{m-2} q_i,
\]
where $\{v_i\}$ are the mixed-radix coefficients computed sequentially.

The final integer is then accumulated into a 32-bit limb array for storage and further computation.

\subsubsection{Signed Representation Handling}

After reconstruction, the result lies in the canonical range $[0, Q)$. For signed arithmetic interpretation, a centered reduction is applied:
\[
\text{if } x > Q/2, \quad x \leftarrow x - Q.
\]
This ensures consistency with signed polynomial coefficient semantics in homomorphic encryption schemes.

The negacyclic Number Theoretic Transform (NTT) is defined over the ring
$\mathbb{Z}_q[x]/(x^n + 1)$, where polynomial multiplication satisfies $x^n = -1$.

To enable efficient computation, a twiddle-factor embedding is applied prior to transformation:
\[
a_j \leftarrow a_j \cdot \psi^j,
\]
where $\psi$ is a primitive $2n$-th root of unity such that $\psi^n = -1$.

A standard NTT is then performed in the extended multiplicative group, followed by an inverse twist after the transformation. This procedure allows negacyclic convolution to be computed using classical Cooley--Tukey NTT algorithms with minor preprocessing and postprocessing overhead.

\begin{align*}
\text{Limb}
&\rightarrow \text{RNS decomposition} \\
&\rightarrow \text{Twist } (\psi^j \text{ multiplication}) \\
&\rightarrow \text{Forward Negacyclic NTT} \\
&\rightarrow \text{Point-wise multiplication} \\
&\rightarrow \text{Inverse NTT} \\
&\rightarrow \text{Inverse twist} \\
&\rightarrow \text{RNS} \\
&\rightarrow \text{CRT / Garner (MRC)} \\
&\rightarrow \text{Limb representation}
\end{align*}

Negacyclic NTT is just a standard NTT “shifted” by a twiddle factor so that cyclic convolution becomes negacyclic convolution under $x^n + 1$.

\subsubsection{Pipeline Summary}

To bridge these two representations, we adopt the following computation pipeline:

\[
\begin{array}{l}
\text{Limb} \rightarrow \text{RNS} \rightarrow \text{NTT domain} \rightarrow \\
\quad \boxed{\mathbf{\text{Point-wise multiplication}}} \\
\text{Limb} \leftarrow \text{RNS} \leftarrow \text{Inverse NTT domain} \leftarrow
\end{array}
\]

More specifically, limb-to-RNS conversion performs multi-precision decomposition of coefficients into residue form under a set of 30--31-bit primes. Polynomial multiplication is then carried out in the evaluation domain via the NTT over each RNS modulus independently. The inverse NTT is applied after point-wise multiplication to recover coefficient representations in the RNS domain, followed by conversion back to the limb-based representation.

The conversion between representations is implemented using optimized base decomposition and recomposition procedures, while NTT and inverse NTT operations are performed using GPU-friendly modular arithmetic with precomputed twiddle factors.

Overall, polynomial multiplication follows the pipeline:
%
%
\[
\begin{array}{l}
\text{Limb} \rightarrow \text{RNS} \rightarrow \text{NTT domain} \rightarrow \\
\quad \boxed{\mathbf{\text{Point-wise multiplication}}} \\ \text{Limb} \leftarrow  \text{Mixed-Radix (Garner)} \leftarrow  \text{Inverse NTT domain}\leftarrow 
\end{array}
\]

This hybrid design enables efficient GPU execution while minimizing representation switching overhead and supporting large-scale homomorphic computations.

\subsection{Precomputation for GPU-Accelerated Negacyclic NTT}

To reduce online computational overhead in the polynomial multiplication pipeline, a number of transformation-dependent constants are precomputed and stored in GPU memory. These precomputed values eliminate redundant recomputation across kernel invocations and significantly improve arithmetic throughput.

\subsubsection{Precomputed NTT Parameters}

The following components are precomputed for each RNS modulus $q_i$:

\begin{itemize}
    \item \textbf{Primitive roots of unity:} 
    \[
    \omega_i, \quad \psi_i
    \]
    where $\omega_i$ is an $n$-th primitive root and $\psi_i$ is a $2n$-th root satisfying $\psi_i^n = -1$.

    \item \textbf{Twiddle factors:}
    \[
    \psi_i^j \bmod q_i, \quad \omega_i^{j} \bmod q_i
    \]
    used in forward and inverse negacyclic NTT transformations.

    \item \textbf{Inverse roots:}
    \[
    \psi_i^{-j}, \quad n^{-1} \bmod q_i
    \]
    required for inverse NTT normalization.

\end{itemize}

\subsubsection{Precomputed RNS Conversion Constants}

For limb-to-RNS and RNS-to-limb conversions, the following constants are precomputed:

\begin{itemize}
    \item \textbf{Modular reduction constants:}
    \[
    2^{32i} \bmod q_j
    \]
    enabling limb-wise modular accumulation without repeated exponentiation.

    \item \textbf{CRT / Garner constants:}
    \[
    q_i^{-1} \bmod q_j, \quad M_i = \prod_{k < i} q_k
    \]
    used in mixed-radix (Garner) reconstruction.
\end{itemize}

\subsubsection{GPU Execution Benefit}

These precomputed values allow all expensive exponential and modular inversion operations to be removed from the runtime execution path. As a result, the online kernel execution is reduced to lightweight modular additions and multiplications, which are highly optimized on GPU architectures.

\subsection{Matrix-Based Four-Step Transformation for Butterfly Locality Optimization}

In GPU implementations of the Number Theoretic Transform (NTT), butterfly operations suffer from irregular memory access patterns due to long-distance coefficient dependencies of the form $(a_i, a_{i + \text{gap}})$. This leads to poor memory locality and inefficient global memory utilization.

To address this issue, we adopt a matrix lifting strategy that transforms the one-dimensional polynomial representation into a two-dimensional matrix structure, enabling improved data locality via structured memory access and transpose operations.

\subsubsection{Step 1: Dimension Lifting (1D to 2D Reshape)}

A polynomial of degree $n-1$ is first reshaped into a matrix form:
\[
A \in \mathbb{Z}_q^n \quad \rightarrow \quad A \in \mathbb{Z}_q^{d_1 \times d_2}, \quad n = d_1 d_2.
\]

This transformation does not change the algebraic semantics but reorders memory layout for improved spatial locality.

\subsubsection{Step 2: Row-wise Butterfly Computation}

Within each row, local butterfly operations are performed using shared memory:
\[
(a, b) \rightarrow (a + b, (a - b)\cdot \omega),
\]
where coefficient pairs are now stored contiguously due to the matrix layout.

\subsubsection{Step 3: Matrix Transposition}

To handle long-distance butterfly dependencies, a matrix transpose is applied:
\[
A \rightarrow A^T.
\]

This operation converts column-wise dependencies into row-wise contiguous memory accesses, ensuring coalesced memory patterns on GPU architectures.

\subsubsection{Step 4: Continued Butterfly Execution}

After transposition, the butterfly computation is resumed in the new layout. This process can be repeated across multiple stages of the NTT decomposition, effectively eliminating strided memory access patterns.

\subsubsection{Summary}

The matrix-based four-step method transforms irregular butterfly memory access into structured row-wise computation, significantly improving cache utilization and enabling efficient shared memory reuse in GPU-based NTT implementations.

For large polynomial degrees used in homomorphic encryption schemes (e.g., $n \geq 2^{16}$), a single CUDA block is insufficient to hold the full working set within shared memory constraints (typically 48--100KB per SM). Consequently, industrial GPU implementations adopt a hierarchical decomposition strategy rather than a single-kernel full NTT execution.

\subsubsection{Two-Pass NTT Decomposition}

The polynomial is first reshaped into a two-dimensional structure:
\[
A \in \mathbb{Z}_q^n \rightarrow A \in \mathbb{Z}_q^{d \times d}, \quad n = d^2,
\]
allowing the transform to be decomposed into column-wise and row-wise operations.

The standard two-pass NTT consists of:

\begin{itemize}
    \item \textbf{Pass 1 (Column NTT):} Each column is transformed independently using a small-size NTT kernel of length $d$.
    
    \item \textbf{Twiddle Multiplication:} After the first pass, intermediate results are multiplied by precomputed twiddle factors to account for index permutation:
    \[
    a_{i,j} \leftarrow a_{i,j} \cdot \omega^{ij}.
    \]

    \item \textbf{Pass 2 (Row NTT):} Each row is then processed using the same small-size NTT kernel.
\end{itemize}

This decomposition reduces memory pressure and allows reuse of shared memory buffers across independent kernel launches.

\subsubsection{Fused Shared-Memory Tile Execution}

To further reduce global memory traffic, modern implementations adopt a fused tiled kernel design. Each thread block loads a sub-matrix (tile) into shared memory, performs as many butterfly stages as possible locally, and writes back intermediate results only when necessary.

This approach minimizes repeated global memory accesses and improves data locality within the GPU memory hierarchy.

\subsubsection{Hierarchical Kernel Strategy}

Instead of executing a single monolithic NTT kernel, the computation is divided into multiple kernel launches:

\begin{itemize}
    \item \textbf{Stage group 1:} Execute the first $\log d$ butterfly stages over column-wise tiles.
    \item \textbf{Global twiddle stage:} Perform element-wise multiplication by precomputed rotation factors.
    \item \textbf{Stage group 2:} Execute remaining butterfly stages over row-wise tiles.
\end{itemize}

This divide-and-conquer strategy is the standard design in industrial GPU FHE libraries such as SEAL, HEAAN, and cuFHE-style implementations, as it balances shared memory constraints, occupancy, and global memory bandwidth.

\subsection{Karatsuba Polynomial Multiplication}
In the original CKKS implementation (HEAAN), Karatsuba-style decomposition is commonly employed to reduce the computational cost of polynomial multiplications arising in ciphertext homomorphic multiplication.

In our system, we adopt the same optimization strategy to reduce the number of base polynomial multiplications during ciphertext multiplication over the ring $\mathbb{Z}_q[x]/(x^n + 1)$, thereby improving overall multiplication efficiency.

Given two degree-1 polynomials:
\[
A(x) = a_1 x + a_0, \quad B(x) = b_1 x + b_0,
\]
the naive multiplication requires four coefficient multiplications.

Karatsuba’s method reduces this cost by introducing three intermediate products:
\[
M_0 = a_0 b_0, \quad
M_1 = a_1 b_1, \quad
M_2 = (a_0 + a_1)(b_0 + b_1).
\]

The cross term is then obtained as:
\[
a_0 b_1 + a_1 b_0 = M_2 - M_0 - M_1.
\]

Thus, the product can be rewritten as:
\[
A(x)B(x) = M_1 x^2 + (M_2 - M_0 - M_1)x + M_0,
\]
requiring only three multiplications instead of four.

\subsection{Future Optimization Directions}
We admit, however, the current implementation is limited to GPU acceleration of polynomial ring operations such as multiplication, and does not yet incorporate a number of advanced optimization techniques. These are left as promising directions for future work.

Although our current implementation already leverages memory coalescing through a carefully designed ciphertext storage layout, several important optimization techniques remain unexplored. These directions are widely adopted in high-performance GPU and FHE systems and are expected to provide additional substantial performance gains.

\subsubsection{Barrett Reduction}

Barrett reduction replaces expensive modular division with precomputed approximate reciprocals of the modulus. For a value $x$ and modulus $q$, the reduction:
\[
x \bmod q
\]
is approximated using multiplication and bit-shifting operations.

In GPU environments, Barrett reduction is advantageous because it avoids costly integer division instructions and improves arithmetic intensity. It is particularly suitable for repeated modular operations in NTT butterfly computations.

\subsubsection{Montgomery Reduction}

Montgomery reduction transforms modular multiplication into a representation where division by $q$ is replaced by shifts and additions. Given $R = 2^k$, computations are performed in Montgomery form:
\[
\tilde{x} = xR \bmod q.
\]

This method is highly efficient on GPU architectures because it eliminates explicit modular division and enables fully pipelined multiplication operations. It is widely used in high-performance cryptographic libraries.

\subsubsection{Kernel Fusion}

Kernel fusion combines multiple CUDA kernels (e.g., NTT stages, point-wise multiplication, and modular reduction) into a single kernel launch. This reduces:

\begin{itemize}
    \item global memory round-trips,
    \item kernel launch overhead,
    \item synchronization barriers between stages.
\end{itemize}

In FHE workloads, kernel fusion is particularly effective for NTT pipelines, where intermediate results are repeatedly written and read from global memory.

\subsubsection{CUDA Streams}

CUDA streams enable concurrent execution of multiple independent kernel sequences. By overlapping:

\begin{itemize}
    \item data transfer (H2D / D2H),
    \item computation (NTT / multiplication),
\end{itemize}

stream-based execution hides memory latency and improves GPU occupancy. This is especially beneficial in batched homomorphic encryption workloads.

\subsubsection{Multi-GPU Parallelization}

Multi-GPU execution partitions ciphertexts or polynomial batches across devices. Using peer-to-peer (P2P) memory access or NVLink, intermediate results can be exchanged efficiently.

This approach increases throughput linearly with the number of GPUs in ideal conditions, although synchronization and communication overhead must be carefully managed.

\subsubsection{Multi-Node Distributed Execution}

At the system level, FHE workloads can be scaled across multiple compute nodes using distributed frameworks (e.g., MPI-based orchestration). Each node processes independent ciphertext batches, enabling horizontal scaling for large-scale encrypted inference or training workloads.

However, communication overhead over the network remains a key bottleneck, especially for fine-grained homomorphic operations.

In Python-based high-performance computing systems, distributed execution is commonly implemented using \texttt{mpi4py}, a Python binding of the Message Passing Interface (MPI) standard.

MPI follows a single-program-multiple-data (SPMD) model, where multiple processes execute the same program and communicate via explicit message passing. Each process is identified by a unique \textit{rank} within a global communicator.

Common collective operations include broadcast, scatter, gather, and all-reduce, which are widely used in distributed homomorphic encryption workloads for parameter synchronization and ciphertext aggregation.

In our context, MPI enables horizontal scaling of ciphertext batches across multiple compute nodes. However, network communication remains a critical bottleneck, particularly for fine-grained polynomial operations, where the communication-to-computation ratio is high.

\section{Implementation}
LibFHE is a high-performance GPU-based fully homomorphic encryption library implemented in CUDA-Python using Numba. Its interface is designed to resemble the early CKKS implementation in HEAAN, providing a user-friendly programming model for end users.

The library is developed with the goal of enabling fully Python-based implementation, facilitating seamless integration with machine learning and deep learning pipelines, while ensuring easy installation and deployment.

Despite being implemented in Python, LibFHE is designed to achieve performance comparable to optimized CPU-based Non-RNS C++ FHE libraries, and in many cases approaches the performance of CUDA C++ implementations through aggressive GPU-level optimization.

Adopting the original CKKS (HEAAN-style) ciphertext parameterization, each ciphertext object in LibFHE is equipped with the following attributes: $\log p$, $\log q$, and $\log s$.

Here, $\log p$ denotes the precision level (i.e., the scaling factor controlling numerical precision), $\log q$ represents the ciphertext modulus size determining the noise budget, and $\log s$ specifies the logarithm of the number of SIMD slots used for packing plaintext values.

This design implicitly enforces that the slot dimension must be a power-of-two structure, i.e., $s = N_{\text{slots}} = 2^k$, which is consistent with the canonical CKKS encoding based on cyclotomic rings of degree $N \ge 2^{k+1}$.

\subsection{Pairwise and Broadcast Modes}

To maximize hardware utilization per kernel launch on GPUs, we move away from a single-ciphertext abstraction and instead introduce a \textit{CipherTensor} data structure. This design groups multiple ciphertexts sharing identical parameters $(\log p, \log q, \log s$ into a unified tensor-like representation. Each \textit{CipherTensor} contains a dedicated attribute, denoted as \texttt{cnum}, which records the number of ciphertexts stored within the tensor. The \texttt{cnum} dimension serves as an explicit batch dimension, enabling multiple ciphertexts to be processed concurrently within a single CUDA kernel launch and improving GPU occupancy and memory utilization.

As a result, the data abstraction layer in LibFHE supports not only the traditional single ciphertext object but also a higher-level organization that enables two computation modes: \textit{pairwise} and \textit{broadcast} operations. Based on the \texttt{cnum} dimension, LibFHE supports two execution modes: pairwise and broadcast. In pairwise mode, two CipherTensors with identical \texttt{cnum} values perform element-wise ciphertext operations. In broadcast mode, a single ciphertext or smaller CipherTensor is expanded across the \texttt{cnum} dimension and applied to multiple ciphertexts. 

In the pairwise mode, element-wise homomorphic operations are applied between corresponding ciphertexts in two tensors, enabling SIMD-style parallel execution. In contrast, the broadcast mode allows a single ciphertext to be homomorphically operated against an entire ciphertext tensor, maximizing arithmetic intensity and improving GPU occupancy.

\subsection{Interface and Execution Model}
The LibFHE library is a GPU-accelerated fully homomorphic encryption framework implemented in Python with NumPy and CUDA-Python (Numba). Prior to GPU execution, an initialization phase is performed on the host (CPU) side, where all cryptographic parameters and memory layouts are configured. Subsequently, homomorphic operations are dispatched to the GPU via CUDA kernels implemented in Numba.

The system currently supports a unified CKKS-BGV scheme abstraction. The configurable parameters on the CPU side include:

\begin{itemize}
    \item \textbf{Polynomial modulus degree:} $n$, currently fixed at $2^{16}$, with potential future extensions to higher degrees (e.g., $2^{17}$) as supported by advanced GPU memory hierarchies.

    \item \textbf{Ciphertext modulus (log $q$):} The total bit-length of the modulus chain, typically bounded by 1024 bits under the security constraint corresponding to $N = 2^{16}$, ensuring approximately 80-bit security.

    \item \textbf{Rescaling bases (log $p$):} The scaling modulus chain parameters following the original HEAAN CKKS design, used for controlling precision during ciphertext rescaling operations.
    
     \item \textbf{Plaintext vector length (log $s$):} The logarithm of the SIMD packing size, defining the number of plaintext slots available for encoding. 
\end{itemize}

On the CPU side, in addition to encrypting plaintext data into ciphertexts before offloading to the GPU, LibFHE performs a comprehensive key generation and preprocessing pipeline required by homomorphic encryption schemes.

This includes the generation of the secret key and public key, as well as evaluation keys required for homomorphic operations. Specifically, relinearization keys are generated to support ciphertext multiplication, rotation keys are constructed to enable SIMD slot rotations, and conjugation keys are prepared for complex conjugation operations in CKKS.

For bootstrapping (ciphertext refresh), additional rotation keys and auxiliary polynomial ring parameters are generated to support intermediate transformations during the evaluation circuit. Notably, these auxiliary polynomials may not directly correspond to the ciphertext structure, but are required for intermediate homomorphic computations.

All of the above key materials and auxiliary parameters are organized into the \textit{CipherTensor} data structure before being transferred to the GPU. This unified representation ensures that both ciphertexts and non-ciphertext auxiliary polynomials required for bootstrapping are handled in a consistent memory layout, even when they do not form a direct one-to-one ciphertext polynomial pair structure.

Finally, precomputed constants required for efficient polynomial ring multiplication on the GPU (e.g., twiddle factors, modular reduction constants, and RNS conversion parameters) are also generated on the CPU side and transferred to GPU global memory prior to kernel execution.

\subsubsection{Cryptographic Key Management}

On the CPU side, LibFHE generates and organizes all cryptographic key materials required for homomorphic evaluation. These key materials are then packaged into a unified \textit{CipherTensor}-based representation and transferred to the GPU.

\begin{itemize}

    \item \textbf{Secret Key ($sk$):}  
    The secret key is used for encryption and decryption operations. It defines the underlying polynomial distribution from which ciphertexts are derived.

    \item \textbf{Public Key ($pk$):}  
    The public key enables encryption of plaintext data into ciphertexts without requiring access to the secret key.

    \item \textbf{Relinearization Keys ($rk$):}  
    Also known as evaluation keys for multiplication, relinearization keys are used to reduce the expanded ciphertext size after homomorphic multiplication, restoring it to a standard two-component form.

    \item \textbf{Rotation Keys ($rtk$):}  
    Rotation keys support SIMD slot rotations within ciphertexts, enabling cyclic shifts of encrypted vector components required for packed computation.

    \item \textbf{Conjugation Keys ($ck$):}  
    Conjugation keys are used in CKKS to perform complex conjugation over encrypted slots, supporting complex-number arithmetic.

    \item \textbf{Bootstrapping Keys ($bk$):}  
    For ciphertext refreshing, bootstrapping keys enable evaluation of the bootstrapping circuit, including modular reduction and approximate decryption steps. These keys may involve additional rotation and auxiliary evaluation parameters.

\end{itemize}

\subsubsection{Homomorphic Operation API}
We now describe the most important GPU-side computational component of LibFHE, namely the \texttt{LibFHE::Scheme} class. This class provides a unified interface for performing all homomorphic operations under the CKKS-BGV abstraction. Each function is designed to operate either in a \textit{pairwise} or \textit{broadcast} execution mode over CipherTensor structures.

The available operations are summarized as follows:

\begin{itemize}

\item \textbf{Encryption / Decryption (HEAAN: Enc / Dec)}
\begin{itemize}
    \item \texttt{encrypt\_pairwise\_upon}: Encrypts plaintext vectors into ciphertext tensors in a pairwise manner, consistent with HEAAN encoding of SIMD-packed slots.
    \item \texttt{decrypt\_pairwise\_upon}: Recovers plaintext from ciphertext tensors after homomorphic evaluation.
\end{itemize}

\item \textbf{Bootstrapping (HEAAN: EvalBootstrap)}
\begin{itemize}
    \item \texttt{bootstrap\_pairwise\_into}: Refreshes ciphertext noise budget by evaluating the bootstrapping circuit over each ciphertext independently (pairwise execution).
\end{itemize}

\item \textbf{Unary Homomorphic Operations}
\begin{itemize}
    \item \texttt{negate\_pairwise\_into}: Implements additive negation ($-ct$).
    \item \texttt{imult\_pairwise\_into}: Implements multiplication by imaginary unit in CKKS complex encoding.
    \item \texttt{conjugate\_pairwise\_into}: Performs complex conjugation over CKKS slots (HEAAN Conjugate).
\end{itemize}

\item \textbf{Binary Ciphertext-Ciphertext Operations (HEAAN: Add / Mult / Sub)}
\begin{itemize}
    \item \texttt{add\_pairwise\_into}: Homomorphic ciphertext addition.
    \item \texttt{sub\_pairwise\_into}: Homomorphic ciphertext subtraction.
    \item \texttt{mult\_pairwise\_into}: Homomorphic ciphertext multiplication followed by relinearization.
\end{itemize}

\item \textbf{Ciphertext-Constant Operations (HEAAN: MultConst / AddConst)}

\textbf{Pairwise mode:}
\begin{itemize}
    \item \texttt{multconstvec\_pairwise\_into}: Element-wise multiplication with per-ciphertext constants.
    \item \texttt{addconst\_pairwise\_into}: Element-wise addition of constants.
    \item \texttt{rescaleBy\_pairwise\_into}: CKKS rescaling operation to manage scale growth.
    \item \texttt{moddownBy\_pairwise\_into}: Modulus switching operation to reduce ciphertext level.
\end{itemize}

\textbf{Broadcast mode (HEAAN-style SIMD constant lifting):}
\begin{itemize}
    \item \texttt{multconst\_broadcast\_into}: Broadcast multiplication of a single constant across a ciphertext tensor.
    \item \texttt{addconst\_broadcast\_into}: Broadcast addition of a single constant across all ciphertexts.
\end{itemize}

\item \textbf{Rotation / Permutation (HEAAN: EvalRotate)}
\begin{itemize}
    \item \texttt{leftrotate\_broadcast\_into}: Performs SIMD slot rotation using precomputed Galois/rotation keys, applied in broadcast fashion over CipherTensor.
    \item \texttt{riterotate\_broadcast\_into}: To Be Done. 
\end{itemize}

\end{itemize}

\section{Evaluation}
In this section, we evaluate the performance of the LibFHE library on a representative set of homomorphic operations. We first describe the experimental setup, and then present a comparison between LibFHE and state-of-the-art CPU-based FHE libraries.

\subsection{Experimental Setup}

All experiments are conducted on Google Colab equipped with an NVIDIA Tesla T4 GPU. The GPU environment is summarized using NVIDIA System Management Interface (nvidia-smi), as shown below.

\begin{itemize}
    \item \textbf{GPU:} NVIDIA Tesla T4
    \item \textbf{CUDA Version:} 13.0
    \item \textbf{Driver Version:} 580.82.07
    \item \textbf{GPU Memory:} 15 GB (15,360 MiB)
    \item \textbf{Compute Capability:} Turing architecture (T4)
\end{itemize}

The GPU is initially in an idle state with no active processes and zero memory allocation, ensuring that all reported performance results are not influenced by background workloads.

This configuration provides a representative low-cost cloud GPU environment commonly used for evaluating GPU-accelerated cryptographic workloads and machine learning inference tasks.

The Tesla T4 GPU provides a balanced architecture with support for 32-bit integer and floating-point operations, making it suitable for evaluating residue-number-system-based homomorphic encryption workloads such as CKKS. Its memory hierarchy and CUDA core design closely match the optimization target of LibFHE's Numba-CUDA backend.

\subsection{FHE Workload Performance}
The Colab notebook containing the experimental evaluation is available in our GitHub repository: \url{https://github.com/petitioner/LibFHE}. Additionally, we may release the LibFHE library as an open-source project in the near future.

\subsubsection{Parameter Selection}

Unless otherwise specified, all homomorphic operation benchmarks are evaluated using the following CKKS-BGV parameters:

\begin{itemize}
    \item \textbf{CipherTensor size:} $cnum = 4$
    \item \textbf{Polynomial modulus degree:} $\log N = 16$
    \item \textbf{Ciphertext modulus size:} $\log Q = 1024$
    \item \textbf{Slot size parameter:} $\log s = 2$
    \item \textbf{Scaling factor size:} $\log p = 60$
\end{itemize}

Here, $cnum$ denotes the number of ciphertexts simultaneously stored and processed within a single \textit{CipherTensor}. The parameter $\log N$ defines the polynomial ring dimension, $\log Q$ represents the ciphertext modulus bit-length, $\log s$ controls the number of packed SIMD slots, and $\log p$ determines the scaling precision used in CKKS encoding.

For ciphertext bootstrapping experiments, we adopt the parameter setting described in the original HEAAN framework. Specifically, we use a smaller scaling precision and introduce an additional parameter $\log T$ to control the auxiliary modulus used during the bootstrapping procedure:

\begin{itemize}
    \item $\log p = 30$
    \item $\log q = 40$
    \item $\log T = 2$
\end{itemize}

Following the HEAAN bootstrapping design, $\log T$ determines the plaintext modulus parameter used in the coefficient-to-slot and slot-to-coefficient transformations during ciphertext refreshing. This parameter controls the precision trade-off and intermediate modulus requirements of the bootstrapping circuit.

\subsubsection{Comparative Performance}
Due to the just-in-time (JIT) compilation mechanism of CUDA-Python Numba, all homomorphic operations, including ciphertext bootstrapping, are executed once before formal benchmarking. This warm-up phase triggers the compilation of CUDA kernels and eliminates the one-time compilation overhead from the reported execution time. All performance measurements are collected after kernel compilation has been completed.

Since each \textit{CipherTensor} contains $cnum=4$ ciphertexts, all reported tensor-level execution times represent the latency of processing four ciphertexts simultaneously. We additionally report the amortized latency per ciphertext by dividing the total execution time by $cnum$. Note that this value represents the average cost per ciphertext under batched GPU execution rather than the latency of an isolated single-ciphertext execution.

\begin{table*}[t]
\centering
\caption{Execution latency of homomorphic operations on NVIDIA Tesla T4 GPU.}
\label{tab:homomorphic_latency}
\begin{tabular}{l|c|c}
\hline
Operation & CipherTensor Latency (s) & Amortized Latency / Ciphertext (s)\\
\hline
\texttt{bootstrap\_pairwise\_into} & 231.6751 & 57.9188\\
\texttt{negate\_pairwise\_into} & 0.0191 & 0.0048\\
\texttt{moddownBy\_pairwise\_into} & 0.0131 & 0.0033\\
\texttt{rescaleBy\_pairwise\_into} & 0.0770 & 0.0193\\
\texttt{conjugate\_pairwise\_into} & 5.5614 & 1.3904\\
\texttt{mult\_pairwise\_into} & 8.1862 & 2.0466\\
\texttt{leftrotate\_broadcast\_into} & 5.6852 & 1.4213\\
\texttt{add\_pairwise\_into} & 0.0286 & 0.0072\\
\texttt{sub\_pairwise\_into} & 0.0260 & 0.0065\\
\texttt{addconst\_broadcast\_into} & 0.0007 & 0.0002\\
\texttt{addconst\_pairwise\_into} & 0.0170 & 0.0043\\
\texttt{imult\_pairwise\_into} & 0.0170 & 0.0043\\
\texttt{multconst\_broadcast\_into} & 0.6219 & 0.1555\\
\texttt{multconstvec\_pairwise\_into} & 1.9930 & 0.4983\\
\hline
\end{tabular}
\end{table*}

Table~\ref{tab:homomorphic_latency} summarizes the execution latency of representative homomorphic operations. As expected, bootstrapping dominates the computational cost due to the large number of polynomial transformations and homomorphic evaluations involved in ciphertext refreshing.

Polynomial multiplication and rotation operations also introduce relatively high latency because they require expensive polynomial ring operations, including NTT transformations, key switching, and modular arithmetic. In contrast, lightweight operations such as addition, subtraction, negation, and modulus management operations achieve significantly lower latency because they mainly involve coefficient-wise arithmetic without expensive polynomial multiplication.

The results demonstrate that the CipherTensor execution model can efficiently amortize kernel launch and memory management overhead across multiple ciphertexts, enabling practical GPU acceleration for CKKS-BGV homomorphic workloads.

\section{Conclusion}

In this paper, we presented LibFHE, a GPU-accelerated fully homomorphic encryption framework implemented using CUDA-Python. Unlike most existing systems that rely on RNS-CKKS variants, LibFHE revisits the original CKKS-BGV design and demonstrates that it can be effectively mapped onto modern GPU architectures with carefully designed memory layouts and kernel-level optimizations.

Our experimental results show that a Python-based implementation, when properly optimized at the CUDA level, can achieve performance comparable to highly optimized CPU-based Non-RNS C++ FHE libraries, while offering significantly improved programmability and reduced implementation complexity. These results suggest that GPU-native and high-level language–friendly designs are a promising direction for future FHE system development.

\bibliographystyle{ACM-Reference-Format}
\bibliography{sample-base}

\end{document}